# Unphysical source of dynamic order for the natural world

R. Herrero [a], J. Farjas [b], F. Pi [c], G. Orriols [c,1]

[a] Departament de Física i Enginyeria Nuclear, Universitat Politècnica de Catalunya, 08222 Terrassa, Spain.

[b] Departament de Física, Campus Montilivi, Universitat de Girona, 17003 Girona, Spain.

[c] Departament de Física, Universitat Autònoma de Barcelona, 08193 Cerdanyola del Vallès, Spain.

## Abstract

The well-defined but intricate course of time evolution exhibited by many naturally occurring phenomena suggests some source of dynamic order sustaining it. In spite of its obviousness as a problem, it has remained absent from the scientific debate due to the lack of candidates for a proper answer. Here we propose a general explanation based on the oscillatory possibilities of the dynamical systems, as demonstrated with a family of differential equations exhibiting self-sustained oscillations of unbounded complexity: complex evolutions by nonlinear mixing of an arbitrarily large number of oscillation modes, in which the different modes describe specific dynamical activities and their combination articulates the interactive conjunction of such activities into the whole functioning. The dynamical scenario is rather generic since it is exclusively based on the reiterative occurrence of the two most standard mechanisms of nonlinear dynamics: the saddle-node and Hopf bifurcations, and its extraordinary richness makes feasible the well-defined occurrence of ordered features over enormously complex dynamical activities as simply arising from the proper structure of dynamical relations among the system components, i.e., without requiring any other physical cause than those involved in such relations.

## Significance

Systems exhibiting rather intricate behaviours over which a well-defined sequence of events is however clearly appreciated are usual in nature. It is particularly obvious when the behaviours exhibit recurrence, like in examples going from the living cell to the celestial motion and passing for the coherent structures of turbulence or the wake-sleep cycle of a brain. The complex dynamical activities occurring within any of these systems should follow some predefined order and from where such a dynamic order can arise is what we try to explain here in generic terms. Our proposal could be useful for guiding the tentative modelling of any of such systems and for facilitating the understanding of how the system attains the coordination of its activities.

Common sense, followed by logical reasoning and the use of mathematical constructs, are the employed rational tools when we, the humans, try to explain our observations in order to achieve some understanding of the world around us and, along the way, the theories of science and philosophy are built up. Consider for instance a living cell and try to imagine the molecular workings sustaining the cell activity up to its division into two. The well-defined sequence of observable intracellular events and, especially, its cyclic recurrence compel our common sense to see obvious that something is there organizing such workings and then the questions of what and where is it immediately arise. The temptation to accept the genome as a director is strong, as it is for the brain in an animal, for the Head in the Physics Department or for some God in the Universe, but logical reasoning quickly put us against the sufficiency of such a kind of explanation simply because any director will need a supervisor and so on. Most importantly, the explanation should cover how the director is orchestrating its supposedly governed elements and, while the philharmonic musicians should be attentive to the conductor, the components of the cell, of the animal or of the universe go clearly under their own steam, only affected by directly interacting neighbours. The relevant question is not who commands, but how the proper ordering of observable events is taking place over the intricate underlying activity. Of course there are elements playing more decisive roles than others but the *unknown thing* we are searching for should actuate both locally, at the range where the physical interactions effectively occur, and on the whole, where the global coordination is manifested.

This article deals with such an *unknown thing* by trying its characterization and by presenting a tentative candidate for it. Being unable to find a proper English name, we use the Catalan expression "*l'entrellat del món*", a euphemism for the underlying reasons ordering the workings of the things of the world, just what we are searching for. The proposed candidate is nothing but a mathematical construct and, then, it does not belong to the physical world.

The same question applies to all levels of life, when passing from cells to organs to whole organisms or when considering the development of a fertilized egg. It points out two interrelated sides of the problem: to what extent the answer could be the same for the different levels and sizes, and of what kind of nature the unknown thing(s) should be. In fact, similar questioning applies to any system exhibiting complex time evolutions in its characteristic properties, independently of it is alive or not. A paradigmatic example is offered by the turbulent behaviour of flowing fluids [1], for which the formation and sustainment of the nested structure of spatiotemporal structures remain pending of explanation since the first inquiry by Leonardo da Vinci (Fig. 1) and in spite of the firm substantiation of the fluid flow physics in the nineteenth century. Leaving apart the different degrees of complexity, the question asking for the source of dynamical order in turbulence is just the same as in the case of the living cell and it applies also to the numerically simulated turbulent behaviours obtained from the Navier-Stokes equation.

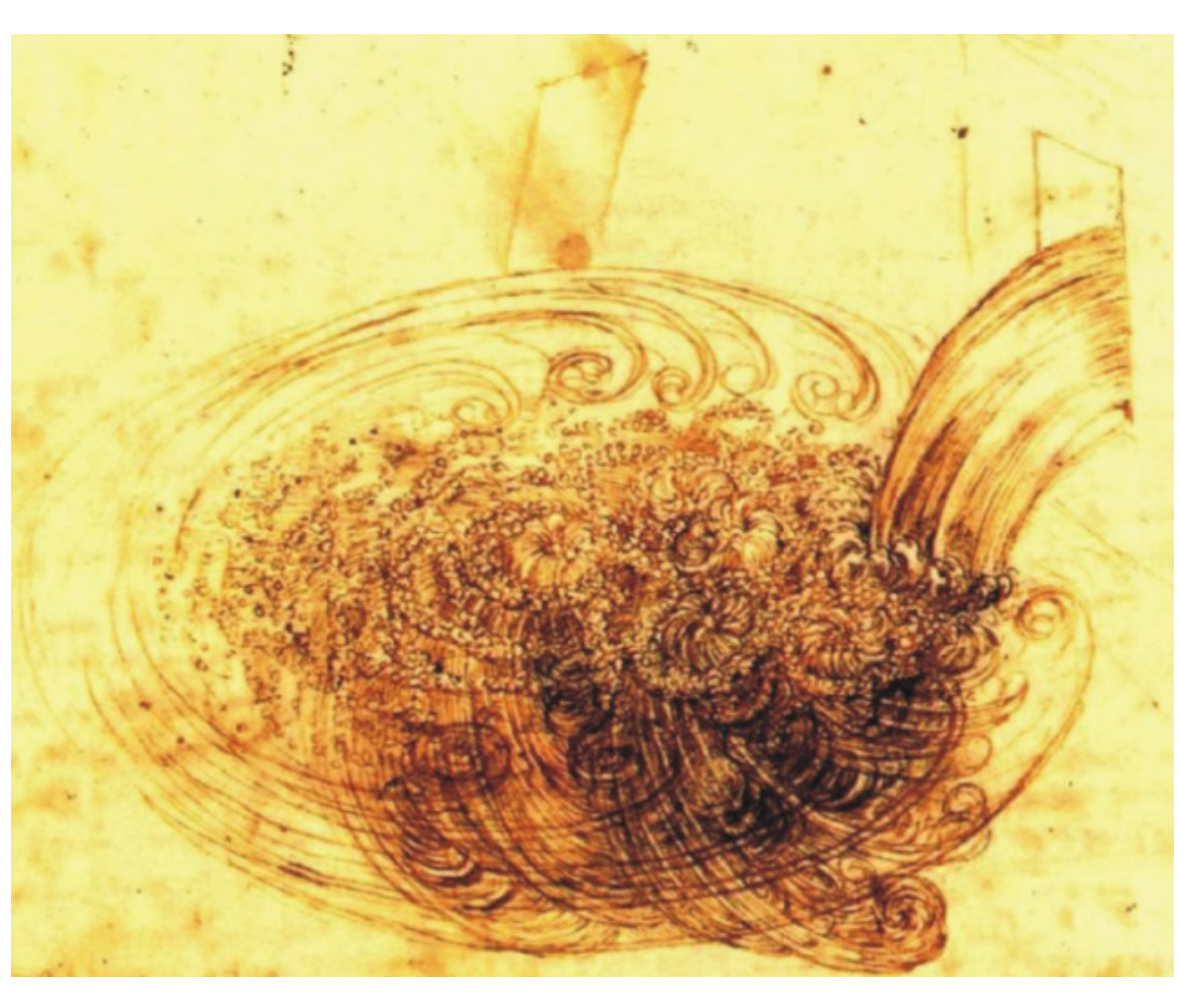

**Figure 1:** Dynamic order in turbulence. Drawing by Leonardo da Vinci excerpted from one of his last studies of flowing water [16]. The pattern of multiple layered whirlpools evolves in time by showing stable and repeatable cycles and this impelled Leonardo to search for the reasons of such a kind of ordered behaviour. An accepted answer is still pending.

Consider now the numerical simulation of Fig. 2 illustrating the asymptotic behaviour of a system of ordinary differential equations. The signal waveform expresses the intermittent combination of oscillation modes of clearly different frequencies and its cyclic recurrence looks practically periodic. Here again, the complex time sequence and, especially, its cyclic recurrence compel the observer to ask about the hidden source of order in the differential equations and in the algorithms simulating their behaviour through a physical computer.

We associate complexity with the behaviours of those systems exhibiting time evolving properties whose sequential details compel the observer to assume unknown sources of order regulating the underlying dynamical effects, as once it was the case for the celestial motion observers. In addition to the orderly manner of functioning, the time evolutions of complexity often manifest structural changes through which the system transforms and usually enriches its dynamical capability.  A tentative theory of complexity should address both features simultaneously by solving their apparent contradiction [2], but here we disregard the structural or creative facet and devote our analysis to the more comprehensible problem of the source of dynamic order, by trying to elucidate from where the observable varieties of ordered functioning can arise. Oddly enough, such a kind of problem has remained absent from the scientific debate, while the structurally creative evolution has been the object of significant efforts tentatively devoted to surpass its contradiction with the second law of thermodynamics [3-5] (see SI Appendix 2 for a distinction between what we call dynamical order and the order/disorder of entropy).

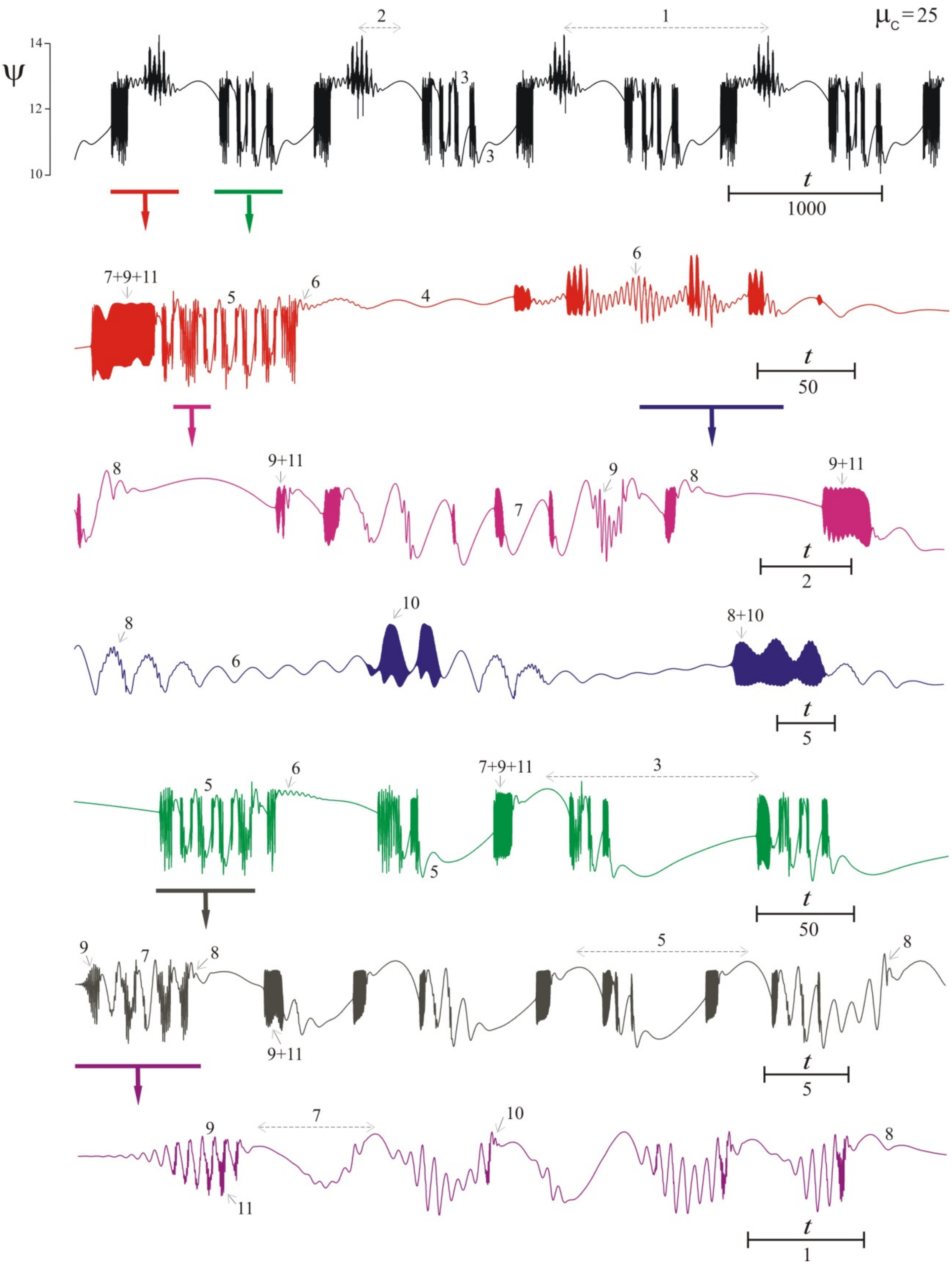

**Figure 2:** Complex and recurrent time evolution derived from a system of 12 differential equations. The evolution shows the intermittent combination of 11 oscillation modes of different frequencies (labelled by numbers). Time scale is dimensionless. For the equations see Fig. 3 and SI Appendix 1. Phase-space representations of the oscillation are shown in Fig. S1.

As today almost standard in the empirical sciences we assume the ontological reductionist view that any observable effect arises exclusively from the involved elementary physical constituents and their fundamental interactions. It has to be distinguished from other reductionisms concerning theories and methods, or from the constructionist hypothesis asserting that any property or phenomenon can be explained by considering elementary details alone. As expressed in the renowned *More Is Different* by Anderson [6], such a constructionist view does often not work and, in general, it is probably unfeasible and perhaps ineffective in providing useful understanding. The scientific explanation has developed through physics, chemistry, biology, and so on, by establishing laws and theories that often seem irreducible to the fundamental levels of physics. The reductionist connection from bottom to top is usually sustained through the notion of emergence, the precise meaning of which, however, is object of philosophical discussion [7].

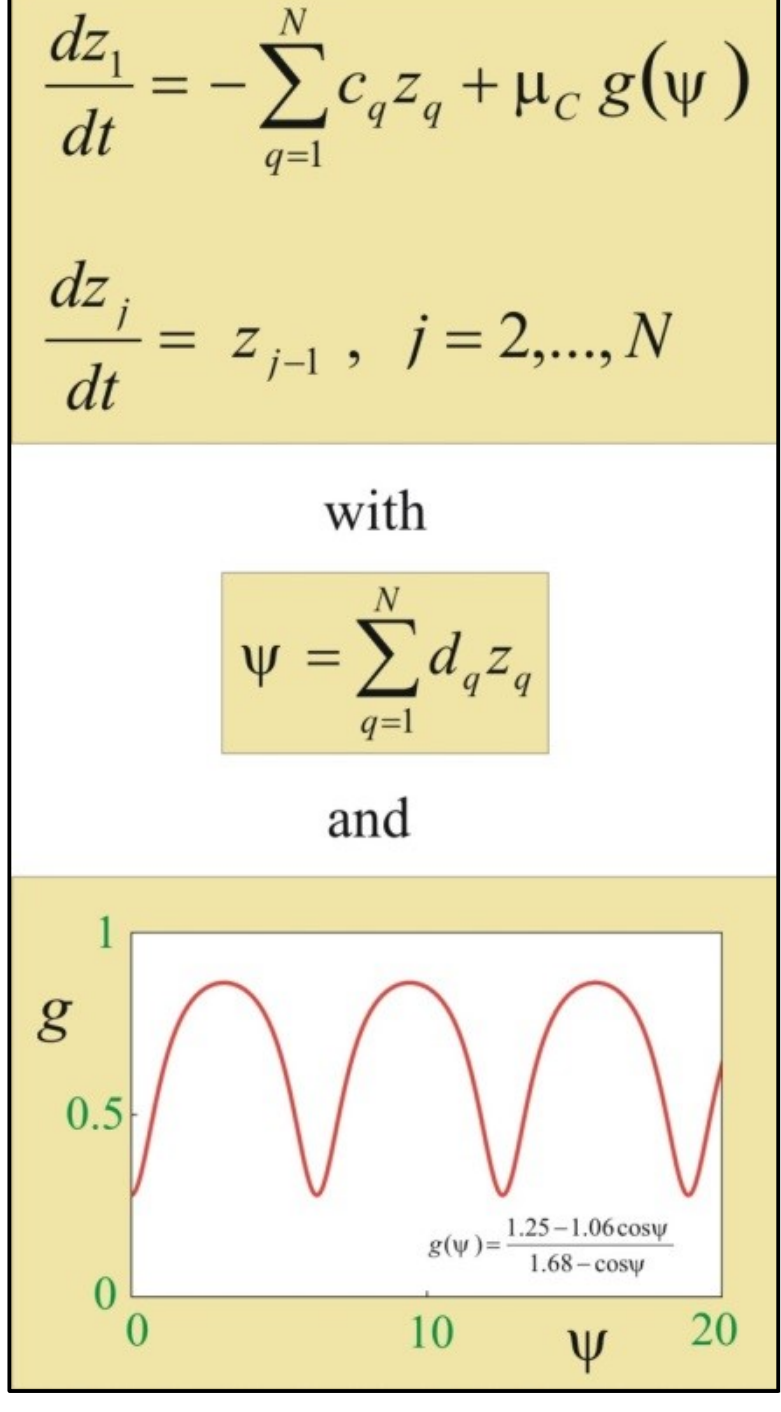

**Figure 3:** System of differential equations used to numerically demonstrate the generalized Landau scenario. The simulations of Figs. 2 and 4 correspond to the values of N, $\mu_C$, $c_q$ and $d_q$ reported in Tables S1 and S2, respectively. The system is a generalization of a model derived to describe a family of physical devices [11]. It admits to be designed [10] by calculating the $c_q$ and $d_q$ coefficients with which a saddle-node pair of fixed points will experience up to N-1 Hopf bifurcations with chosen values for their frequencies. More details in SI Appendix 1.

The physical mechanisms, at either fundamental or emergent levels, are intrinsically scale dependent. Thus, by taking into account the extensive variety of scales for which a source of dynamical order seems necessary, a huge number of different types of sources would be required if they would be physically based. Such a proliferation looks implausible, especially because no one of them has been observed. An alternative view is offered by the field of nonlinear dynamics [8,9] through its most basic message stating that the qualitative behaviour of a system arises from the structure of dynamical relationships, independently of the concrete nature of the effects actually sustaining such relations. Thus, by tentatively identifying the workings of real systems with those of mathematical dynamical systems, our problem transforms into the two-step question:

1. Are there generic mechanisms in nonlinear dynamics powerful enough to generate so complex behaviours as needed for complexity?
2. What determines the behaviour of the corresponding set of differential equations?

The "What" of the second point refers to the source of order, it is necessarily related to the ensemble of dynamical relations expressed in the given system of equations, since there is nothing more in the equations, and the proper answer will be found by analysing how such relations determine the system behaviour. The independence of the concrete physical nature of the actual relations will provide the hypothetical source of order with full genericity, permitting its coherent operation at the different scales, levels and circumstances, and, on the other hand, will make it compatible with the reductionist view of causal influences from bottom features exclusively.

At this point we need to justify why such a view is not pervading the stream of science and it has to do with the lack of a proper answer to the first of the questions above: the mainstream of nonlinear dynamics lacks knowledge about generic scenarios sustaining the dynamically organized interplay of high numbers of degrees of freedom with a growing amount of differentiated activities, as it seems to be the case in complexity behaviours. Chaos, the main object of research during years, is associated with low-dimensional processes and there is no indication among its features of an effective way for the coordinated accumulation of dynamical activities. Such a kind of way has not been found in studies of either high-dimensional systems based on coupled discrete sets or continuous spatially-extended structures. However, such a kind of way exists [10,11,2] and it is our goal to spread knowledge of the corresponding dynamical scenario and to show how its extraordinary possibilities make it suitable as underlying basis for the ordered functioning of complexity.

**Generalized Landau scenario**

The second most basic message from nonlinear dynamics states that dynamical activity is synonymous of oscillatory activity. The time evolution of variable properties implies either increases or decreases of their values and, since the exhibition of monotonic variations seems irrelevant as an activity, the sole possibility is the oscillatory alternation. Any complex dynamical activity, presumably including that occurring in a living cell as a whole, should be nothing but some kind of complex oscillation and then the question of how complex the dynamical oscillations may be compulsorily rises. Nevertheless, concerning the generation of self-sustained oscillations, a significant drawback arises in nonlinear dynamics from the almost unanimous convincement that the exclusive generic way for combining oscillations is through the torus bifurcation, and from the deluding verification that such a way does not work in general [12,13]. High-dimensional tori are relevant in Hamiltonian systems but their fragility enhances with dissipation, they easily break by leading up to chaos, and the consequence is the practical absence of tori of order higher than two in non-conservative autonomous systems. Under this dominant view, the intuitively convincing idea that complex oscillations could be achieved by combining more and more oscillations, as proposed by Landau to explain the transition to turbulence [14], looks unachievable in the form of a self-oscillating behaviour.

The research of complex dynamics has focused its attention onto chaos, a kind of oscillatory behaviour exhibited by systems located (in the space of dynamical systems) in the middle of extraordinarily dense accumulations of codimension-one bifurcations of different types, through which indefinitely high numbers of nearby periodic orbits can coexist in the same region of the phase space [8,9]. If one of the orbits is stable, the system state describes the corresponding periodic evolution but typically all of them are saddles and then the state evolves irregularly as describing close passages to an indefinite succession of such periodic orbits. However, chaos is achievable in three-dimensional phase spaces, where the great diversity of periodic orbits should necessarily describe very similar dynamical activities in the physical space. The rather complex dynamics of chaos does not correspond to what seems to be involved in the behaviours of complexity: an arbitrarily large number of interacting properties coordinately sustaining a great diversity of dynamical activities. This is just what is offered by the *generalized Landau scenario* [2,10,11], in which a succession of oscillatory modes associated with different degrees of freedom combine ones within the others without requiring invariant tori and without a limiting reason in their number other than the phase space dimension (see SI Appendix 3 for an overview of the scenario and its difference with chaos).

Figure 4 and Figures S2-S4 illustrate the mechanisms of oscillatory mixing in systems of successively increasing dimension N and suggest their extrapolation to higher dimensions. In the case N=12 of Fig. 2 eleven oscillation modes have emerged in a twelve-dimensional phase space through a succession of Hopf bifurcations, six of them occurring in an initially stable fixed point up to exhaust its stable manifold and other five in a saddle fixed point initially having one unstable dimension connecting it to the first fixed point. The observed attractor derives from the first periodic orbit born from the stable fixed point, while the rest of periodic orbits emerged either from this point or from the saddle point are saddles with a network of interconnections among them based on their unstable manifolds and with all of them connected to the attractor. The oscillatory motions do not remain on the respective periodic orbits but extend towards significant phase space regions around their unstable manifolds and a robust kind of oscillatory mixing occurs that manifest in the periodic orbits themselves by incorporating localized influences of other periodic orbits. The incorporative mixing takes place without requiring any bifurcation of the orbit but simply through the gradual intertwinement of trajectories around the incoming manifold. In particular, the attractor incorporates localized contributions of all the oscillation modes, as manifested in the time evolution by the nested structure of intermittent bursts.

On the basis of a saddle-node pair of fixed points, the scenario can originate up to N-1 different oscillation modes, with no limits to the N value, and additional fixed points can participate provided that all of them belong to the same basin of attraction, i.e., one attractive fixed point surrounded by a cloud of saddle points connected among them and to the attractive one. Every fixed point allows for the contribution of all the degrees of freedom, those associated with its initially stable dimensions can sustain the appearance of oscillation modes, one for each pair of dimensions, and those associated with its initially unstable dimensions propagate the influence of these modes towards other fixed points to sustain the oscillatory mixing with the modes emerged from them.

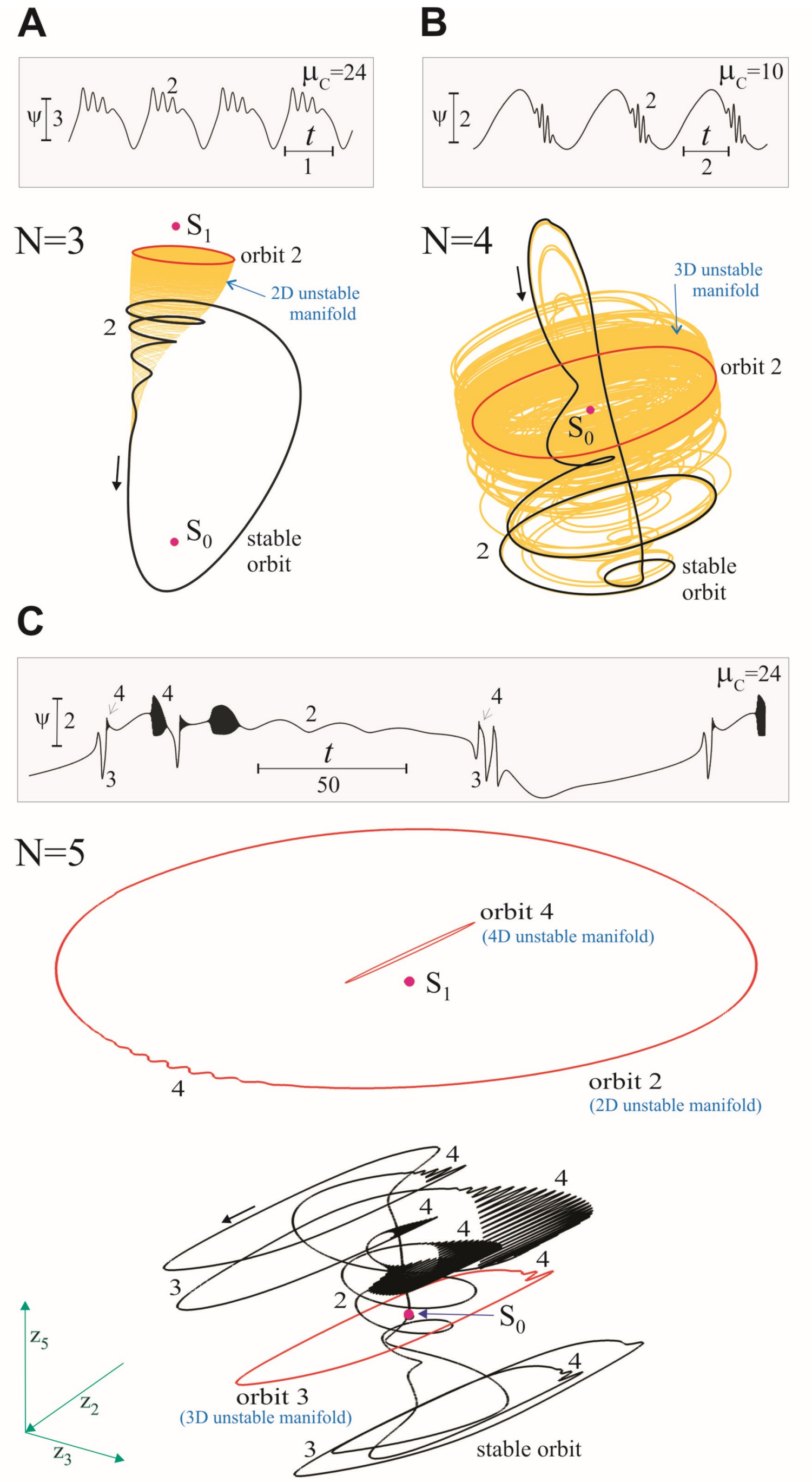

**Figure 4:** Mixing of oscillation modes in systems of successively increasing dimension N. (A) Mixing of two modes appeared from a saddle-node pair of fixed points through the respective Hopf bifurcations. (B) Mixing of two modes emerged in successive Hopf bifurcations of the same fixed point. (C) Mixing among four modes emerged from a saddle-node pair of fixed points experiencing both two successive Hopf bifurcations. In each case, it is shown the time evolution of the stable orbit and a phase space representation of the various orbits, ordered by numbers according to their frequency. $S_j$ denotes a fixed point with j initially unstable dimensions. The first orbit emerged from $S_0$ is stable (in black) while the rest are saddles (in red) extending their oscillation along the unstable manifold towards other orbits. In the three cases, the stable orbit has not suffered any bifurcation from its origin to the represented situation, while mode mixing has occurred through intertwinement of trajectories. See additional details of (C) in Figs. S2-S4. The parameters of the simulations are in Table S2.

The oscillation modes extend their influences over the phase space by maintaining both the frequency and the orientation of the corresponding periodic orbits. This relevant fact means that each mode describes a well-defined activity in which the different variables participate to a greater or lesser extent according to the oscillation plane projection on the respective axes. The mode mixing expresses how a given oscillatory activity affecting certain variables modulates the actuation of another activity, usually faster and affecting another set of variables, and so on along the chain of influences within the structure of unstable manifolds of the periodic orbits. The number of oscillation modes represents the number of characteristic activities the scenario is combining and such a number can indefinitely grow with the appropriate involvement of more dynamically relevant properties. The potential ways of mixing the scenario can sustain allows us to imagine extraordinarily complex sequences of combined dynamical activities covering a multitude of disparate time domains but describing a well-defined time evolution. Just what we need: unlimited possibilities of dynamical enrichment under a robust basis of ordered functioning.

One might suspect that the systems exhibiting such complex oscillations are rare and then irrelevant for practical purposes but this is not the case. In the space of dynamical systems (see SI Appendix 3), the generalized Landau scenario comes forth through successive crossings of two kinds of codimension-one bifurcation surfaces, the saddle-node and Hopf bifurcations, while the oscillatory mixing happens without requiring any bifurcation. Generally speaking, the scenario develops, and reversely dismantles, as a gentle process associated with the gradual intertwinement of trajectories around the unstable manifolds of the periodic orbits and with the successive incorporation of other fixed points and new periodic orbits. The regions of oscillatory systems extend in continuity towards higher dimensions, without any disruption at the crossing of additional saddle-node and Hopf bifurcation surfaces or when crossing the densely accumulated bifurcations of chaos. Only certain global bifurcations of homoclinic nature can destroy the attractor but without altering the oscillatory mixing scenario that then will contain transient trajectories eventually evolving towards another basin of attraction.

**“L’entrellat del món”**

In a hypothetical world whose things behave like the mathematical dynamical systems, the properties involved in the interrelations sustaining the dynamical activity and the interrelations themselves should remain well-defined along the time. The creative facet of complexity should not occur in such a world and the unique variation will be on the magnitudes quantifying the involved properties. Notice, however, that in our actual circumstance of observers, any attempt to describe a given piece of reality like a dynamical system inevitably commits strong simplifications in the considered details and the observed structural transformations could result from the unexpected actuation of omitted details [2].

To what extent the physical world where we are thinking can be associated with such a hypothetical world is debatable from multiple points of view but, since the differential calculus invention, the scientific attempts to tackle any sort of dynamical phenomena, either at fundamental or particular levels, have been indefectibly based on some kind of differential equations. The successes achieved and the lack of alternatives suggest a close connection between the source of order of the differential equations and that of the things of the physical world. Concerning complexity behaviours, such an association acquires sense through the dynamical possibilities of the generalized Landau scenario and then we proceed with the analysis by tentatively seeing the complex dynamical activities of real things as based on the oscillatory scenario. This view agrees with the documented ubiquity of rhythms, cycles and oscillatory bursting in a wide variety of natural phenomena covering all the spatial scales.

Since the exclusive contents of the differential equations are the relations among the variables and their time rates of change, it is in the ensemble of such interrelations where the source of order must reside, and this would tentatively apply also to the causal interrelations of the physical system to which we associate the given set of equations. Nonetheless, a large variety of transformations of the mathematical system yield new sets of variables sustaining profoundly different ensembles of interrelations while their behaviours remain qualitatively equivalent. The source of order as such should remain unchanged under these transformations but no unchanging features can be appreciated at the relational level, where the connection to the physical world lies, and we must move to the abstract phase space to search for the underlying reasons defining the system behaviour. Such reasons are expressed in the structure of invariant sets of the system: the existing limit sets, mainly fixed points and periodic orbits, their stability features and how their stable and unstable invariant manifolds expand through the phase space by connecting ones with others. The qualitative features of these invariant sets delineate those of the rest of phase space trajectories and they certainly remain under the system transformation.

The reasons we are searching for should be those defining both the possibilities and the restrictions in the appearance and assembling of such invariant sets throughout the oscillatory scenario unfolding. Some of the reasons emanate from the intrinsic determinism of the differential equations description [15] as manifested by the existence and uniqueness theorems and the consequent constraint of no-intersection of phase space trajectories, and by the continuity theorems concerning the spread of trajectories and how the successively appearing limit sets and their invariant manifolds should qualitatively be in relation to the previous ones. Reasons of a different kind arise from the topological constraints delimitating the oscillatory mixing pathways, which, as anticipated in the intertwinement of periodic orbits and their invariant manifolds, become rather cumbersome with increasing the number of orbits and the manifold dimensions. Globally, these generic reasons prefigure the repertoire of permitted oscillatory behaviours and, at the same time, constitute the source of order from which each time evolution will occur in a well-definite manner.

The temporal changes exhibited by a physical system result from the causal interrelations among its dynamically effective properties. Appropriate ensembles of such relations can sustain the rather complex behaviours of the generalized Landau scenario, with the system details defining the concrete scenario development and its physical realization through the actual activities (properties and frequencies) associated with the combined oscillatory modes. Nevertheless, the explanation of how the ensemble of physical interrelations defines the system future and of why this future is as it will be resides at the abstract level of the phase-space generic reasons. In short, “l’entrellat del món” does not exist at all. It merely works.

## Acknowledgments

The writing of the paper in its actual form was suggested and stimulated by R. Corbalán after a talk by one of the authors. Fruitful discussions with J.M. Masjuan helped to improve the manuscript. Financial support by MINECO of Spain under Grant FIS2014-57460P is acknowledged.

## Footnotes

[1] To whom correspondence should be addressed. Email: gaspar.orriols@uab.cat.

Author contributions: All authors collaborated in developing the main concepts and performing the numerical simulations and their analysis. G.O. wrote the manuscript, with feedback from all authors.

The authors declare no conflict of interest.

## Supporting Information for

# Unphysical source of dynamic order for the natural world

R. Herrero[1], J. Farjas[2], F. Pi[3], G. Orriols[3]

[1]Departament de Física i Enginyeria Nuclear, Universitat Politècnica de Catalunya, Terrassa, Spain.

[2]Departament de Física, Universitat de Girona, Girona, Spain.

[3]Departament de Física, Universitat Autònoma de Barcelona, Cerdanyola del Vallès, Spain.

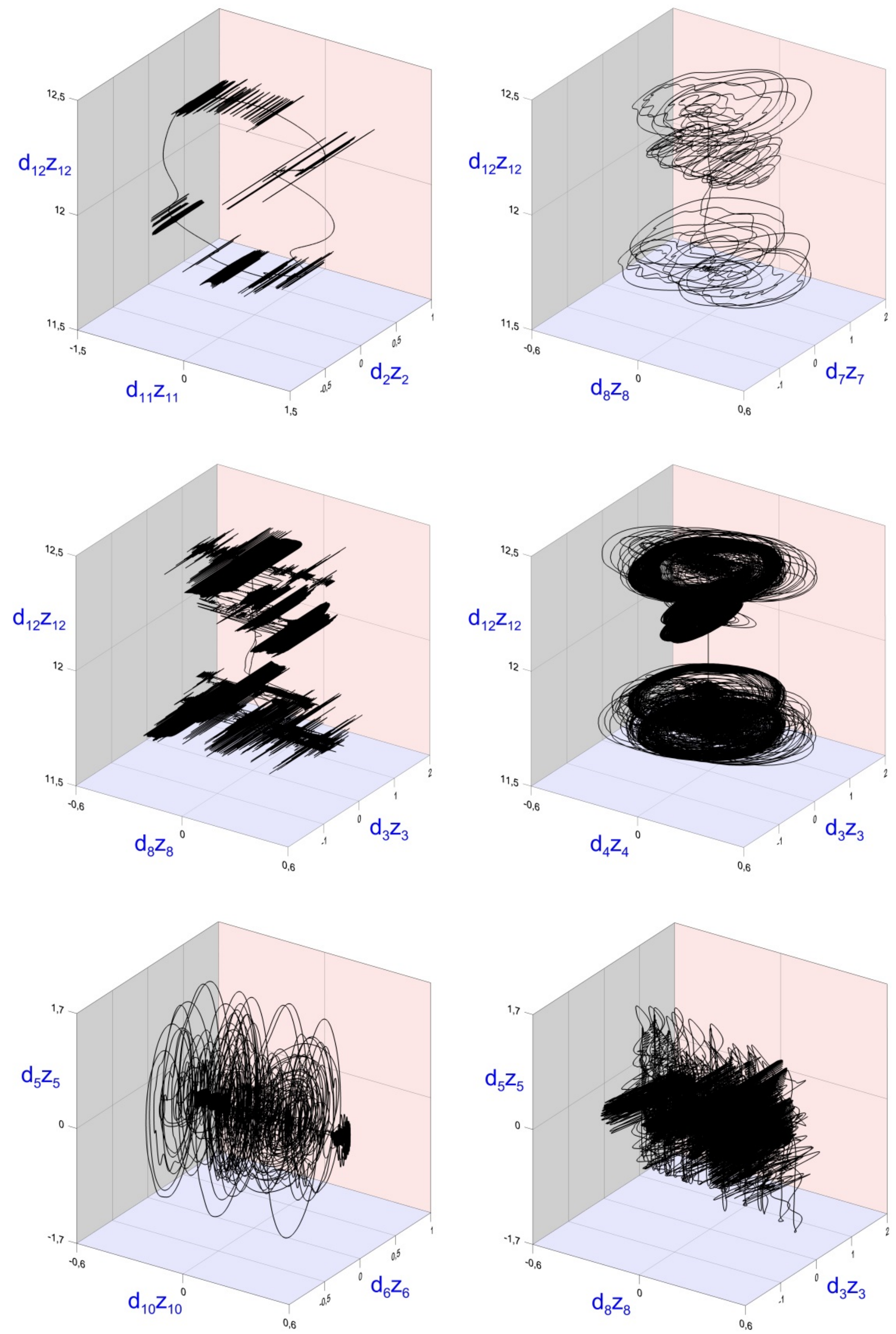

**Figure S1:** Phase space representations of the signal of Fig. 2, showing one of the cycles projected on different three-dimensional subspaces of the twelve-dimensional phase space. The saddle-node pair of involved fixed points have $d_{12} z_{12}$ equal to 13.41 and 11.88, respectively, with the rest of coordinates equal to 0, so that in the projections lacking the $z_{12}$ axis both fixed points appear superposed on the origin. In twelve dimensions, the orbit structure is not as intricate as it appears in the projections since each one of the eleven oscillation modes will describe orbits of peculiar shape and orientation, in correspondence with the eleven periodic orbits, six of them located around the node point and five around the saddle.

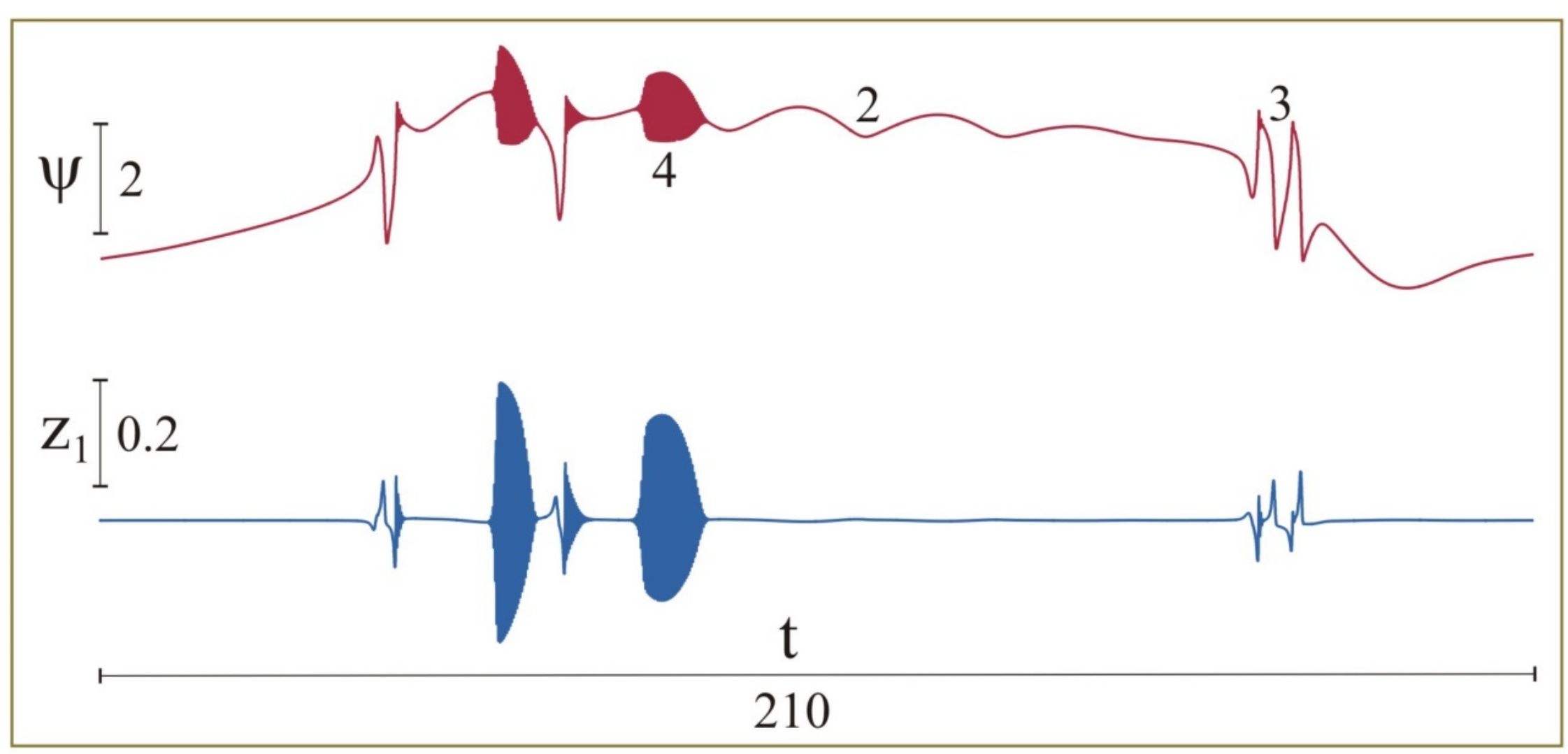

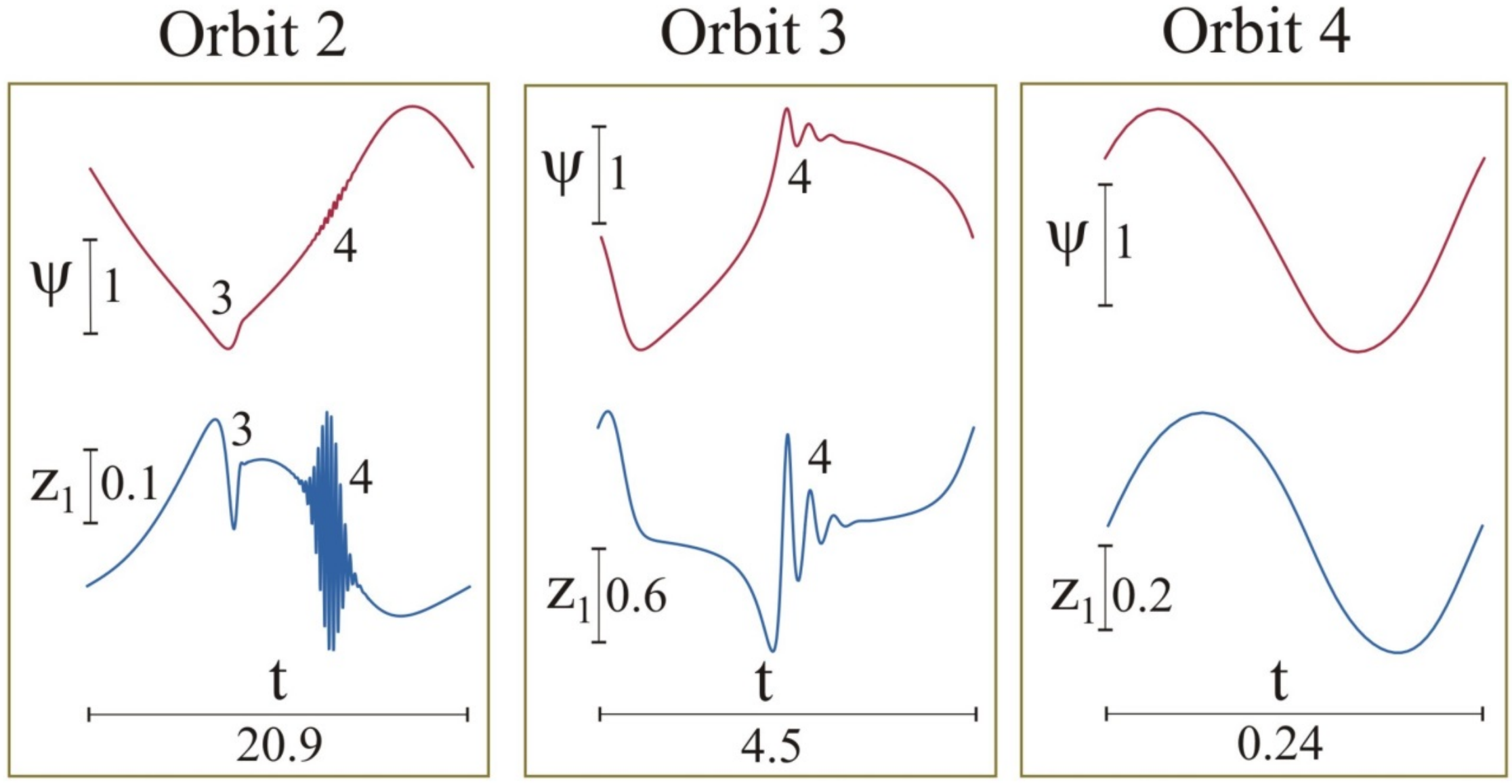

**Figure S2:** Time evolutions of the four periodic orbits of Fig. 4C. The signals describe just one period of the orbits for two variables. Notice the influence of higher frequency modes on the orbits of lower frequency. Particularly relevant is the influence of mode 3, born from the node point, on the orbit 2, born from the saddle point.

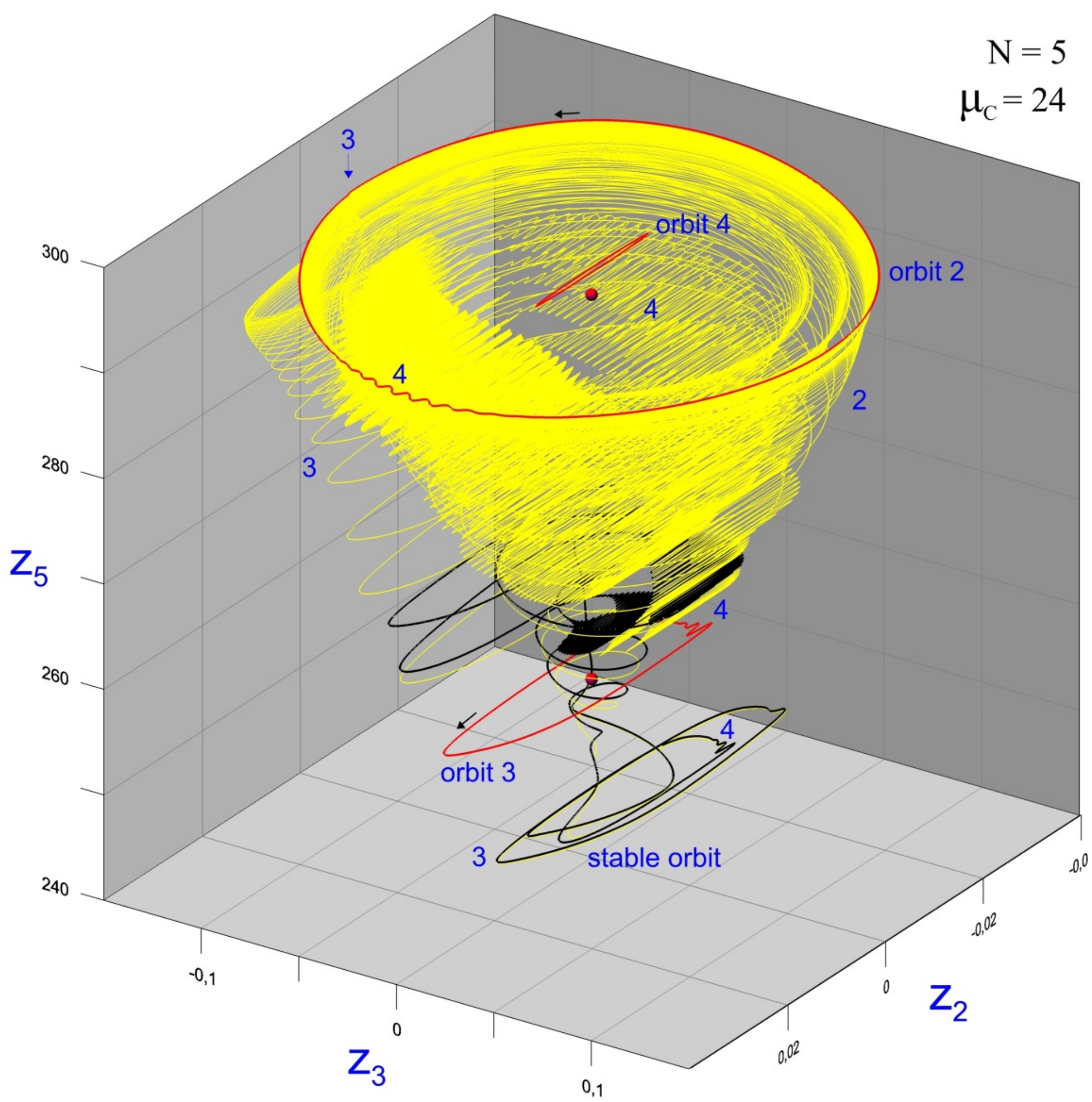

**Figure S3:** Unstable manifold of orbit 2. The same as in Fig. 4C but with a transient trajectory (in yellow) belonging to the two-dimensional unstable manifold of orbit 2. Twenty of such trajectories initiated from different points along the periodic orbit have been computed, all of them look rather similar and they simply fill the two-dimensional surface by maintaining the structure that is better appreciated with a single trajectory. In the way to the attractor, the trajectory describes a combinatory sequence of the oscillation modes associated with the three saddle periodic orbits, pointing out how mode mixing strongly affects large phase space zones, in addition to the periodic orbits themselves. The unstable manifold of orbit 3 is three-dimensional and goes exclusively to the stable orbit, i.e., there is no unstable submanifold of orbit 3 going towards orbit 2. This means that the mixing mechanism is more general than what is contained in our description. Orbit 4 has born at $\mu_C = 21.83$ with a four-dimensional unstable manifold, but at $\mu_C = 23.58$ has stabilized in two dimensions by experiencing a subcritical torus bifurcation with the secondary frequency just equal to that of orbit 2. Then, at $\mu_C = 24$ there should be the two-frequency limit set(s) associated with that torus which would have the four-dimensional unstable manifold involved in the mode mixing with orbit 2 and towards the attractor. Orbit 4 remains with a two-dimensional unstable manifold that goes down towards the attractor without being involved in any mixing. The orbits 2 and 3 have not bifurcated from their origin to $\mu_C = 24$.

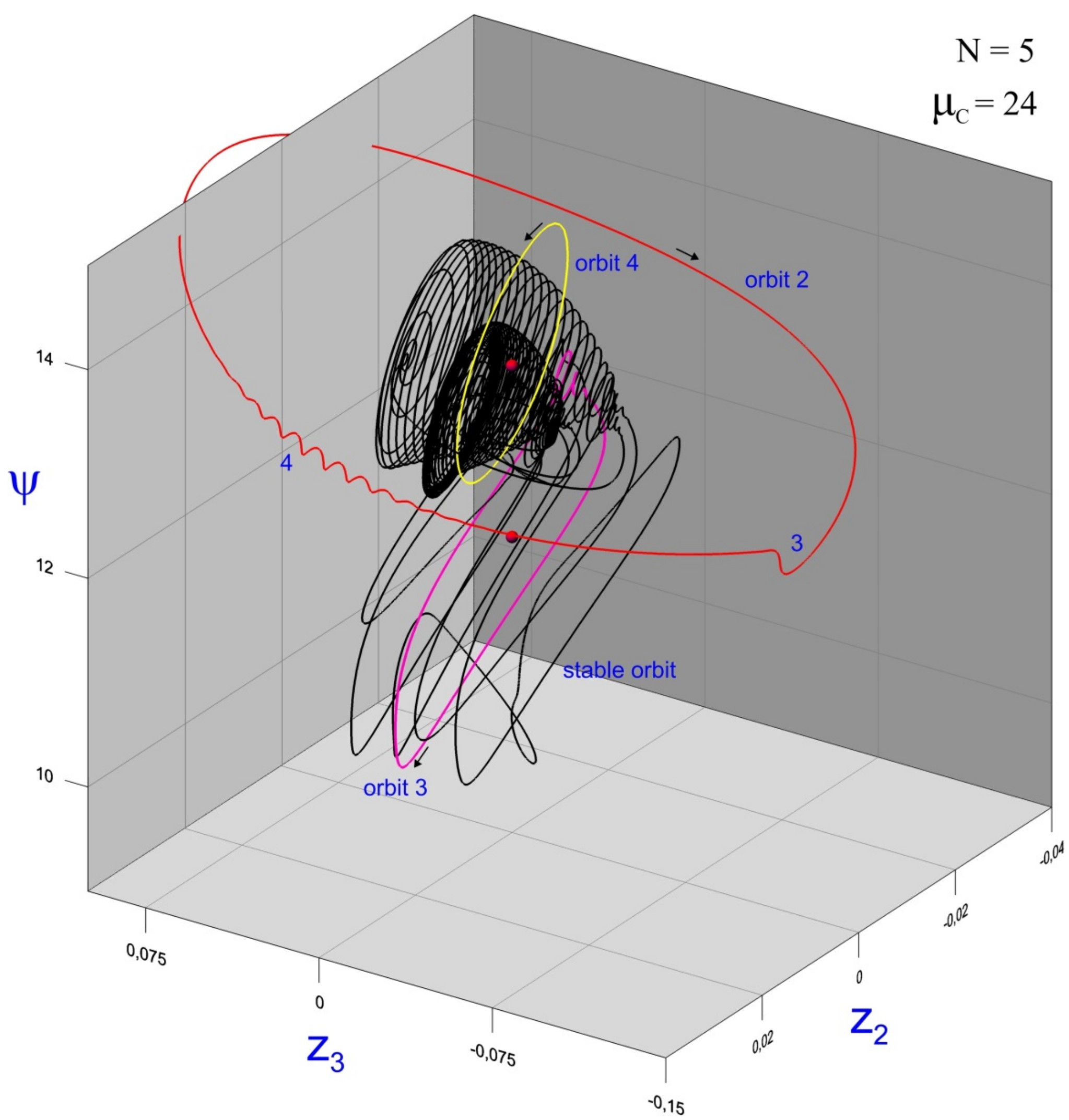

**Figure S4:** Different three-dimensional projection of the orbits of Fig. 4C, with the vertical scale changed to $\psi$ instead of $z_5$ and with the $z_3$ axis inverted. Unlike $z_5$, the variable $\psi$ is sensitive to all the oscillation modes and then reinforces the higher frequency modes. In particular, the influence of mode 3 on orbit 2 is now clearly appreciated while it appears like a minute fold in Fig. 4C. In any case, it is worth recalling that the representation is a three-dimensional projection of a five-dimensional phase space.

**Table S1.** Parameters used in the simulation of Fig. 2

| N = 12 ; $\mu_C$ = 25 | |
|---|---|
| $c_q$ | $d_q$ |
| 1000 | -126.4 |
| 2220 $10^2$ | 2193 $10^1$ |
| 4207 $10^4$ | -5831 $10^3$ |
| 7942 $10^5$ | 9956 $10^4$ |
| 1798 $10^7$ | -2645 $10^6$ |
| 3015 $10^7$ | 4153 $10^6$ |
| 7727 $10^7$ | -1194 $10^7$ |
| 1171 $10^7$ | 1685 $10^6$ |
| 3281 $10^6$ | -5457 $10^5$ |
| 3954 $10^4$ | 6512 $10^3$ |
| 1543 $10^3$ | -2653 $10^2$ |
| 405.2 | 405.2 |

These $c_q$ and $d_q$ values have been obtained by rounding to four digits those calculated by designing a system of equations (Fig. 3 and SI Appendix 1) with N=12 and Hopf frequencies: 0.006, 0.022, 0.06, 0.21, 0.63, 2.1, 6.3, 21, 63, 209, 628, which roughly correspond to the periods: 1047, 285, 105, 30, 10, 3, 1, 0.3, 0.1, 0.03, 0.01, in arbitrary time units.

**Table S2.** Parameters used in the simulations of Fig. 4

| (**A**) N = 3 ; $\mu_C$ = 24 | | (**B**) N = 4 ; $\mu_C$ = 10 | | (**C**) N = 5 ; $\mu_C$ = 24 | |
|---|---|---|---|---|---|
| $c_q$ | $d_q$ | $c_q$ | $d_q$ | $c_q$ | $d_q$ |
| 30 | 4 | 50 | -16 | 20 | 2.9 |
| 385 | -75 | 440 | 66 | 323 | -50 |
| 343 | 343 | 480 | -200 | 59 | 10 |
| | | 360 | 360 | 26 | -5.1 |
| | | | | 0.046 | 0.046 |

In case (**C**), the coefficients $c_q$ and $d_q$ have been obtained by rounding the values calculated by designing a system of equations (Fig. 3 and SI Appendix 1) with N= 5 and Hopf frequencies: 0.05, 0.3, 1.8, 26, which correspond to the periods: 125.7, 20.9, 3.5, 0.24.

## SI Appendices

### 1. Differential equations for the numerical simulations.

A very general description of the $N$-dimensional dynamical systems is

$$\frac{dx}{dt} = Ax + \sum_{j=1}^{m} b_j\, f_j(x,\mu)\,, \tag{S1}$$

where $x \in \Re^N$ is the vector state, $A$ is a constant $N$x$N$ matrix, $b_j$ are constant $N$-vectors, $f_j$ are scalar-valued functions nonlinear in $x$, $\mu$ describes constant parameters involved in the nonlinear functions, and the $m \leq N$ components $b_j f_j$ are linearly independent. Under appropriate nonlinearities, the system (S1) may possess $m$-dimensional arrays of fixed points and a basin of attraction can involve up to $3^m$-1 saddle fixed points of different types in addition to the attracting one [2].

For $m$ = 1 and provided that the matrix ($b_1, Ab_1, A^2 b_1, \ldots, A^{N-1} b_1$) has rank equal to $N$, system (S1) can be linearly transformed in a standard form like

$$\begin{aligned} \frac{dz_1}{dt} &= -\sum_{q=1}^{N} c_q z_q + f_1(z,\mu), \\ \frac{dz_j}{dt} &= z_{j-1}\,, \quad j = 2,\ldots,N, \end{aligned} \tag{S2}$$

where $z$ is the new vector state and $z_q$ its components. The fixed points would appear located on the $z_N$ axis. The design of the system [10] is facilitated by considering nonlinear functions of a single variable in the form

$$f_1(z,\mu) = \mu_C\, g(\psi,\mu)\,, \tag{S3}$$

with

$$\psi = \sum_{q=1}^{N} d_q z_q\,, \tag{S4}$$

and where $\mu_C$ will be taken as a control parameter. The sole requirement on $g(\psi)$ is that it should describe some sort of hump to allow for the coexistence of more than one fixed point, while its actual expression has a secondary influence on the oscillatory behaviour. The coefficients $c_q$ and $d_q$ are the relevant parameters since their values determine the occurrence of Hopf bifurcations in the several fixed points. Under proper choice of them, the corresponding family of systems defined by increasing $\mu_C$ penetrates well into the region of oscillatory systems up to reach the nonlinear combination of $N$-1 oscillation modes, as originated from a saddle-node pair of fixed points. Starting from one of such $N$-dimensional families, it is easy to appreciate the robustness of the oscillatory behaviour and its noncritical localization in the space of dynamical systems by verifying how slightly the behaviour varies when the values of the $c_q$ and $d_q$ coefficients are gradually modified or when the $g(\psi)$ function is changed. It is also feasible to design families of higher dimension with additional oscillation modes by maintaining the main oscillatory features of the starting one.

The simulations reported in Figs. 2 and 4 correspond to Eqs. (S2-S4) with the $c_q$ and $d_q$ values specified in Tables S1 and S2, respectively, and with the periodic nonlinear function

$$g(\psi)=\frac{1.25-1.06\cos\psi}{1.68-\cos\psi}\ , \tag{S5}$$

that describes the interferometric Airy function of the family of physical devices through which the oscillatory scenario was discovered [11]. A graphical representation of Eq. (S5) is shown in Fig. 3. The function periodicity allows for a multiplicity of coexisting fixed points but the scenario usually develops from one stable fixed point and one of the two saddle points bordering the corresponding basin of attraction. In the design process, different enough frequencies for the several oscillation modes have been chosen in order to facilitate their identification on the time evolution signals. The nonlinear mode mixing works also for more similar frequencies but the waveform structures will become gradually blurred up to look like the irregular signals of chaos.

In the figures, the numeric labels denote the oscillation modes ordered according to their frequencies. On the other hand, in the design process, the Hopf bifurcations have been alternatively chosen such that the modes with odd labels have emerged from the node point and those with even label from the saddle point, with the exception of Fig. 4b where the Hopf bifurcation of the saddle will occur at a higher $\mu_C$ value and the corresponding mode is even absent in the mixing on the attractor. Note, in particular, how the abundant multi-frequency bursts of Fig. 2 correspond to modes of either the node or the saddle fixed point.

The peculiar structure of system (S2), with the simple differentiation relation between successive $z_j$ variables, implies that the relative presence of the oscillation modes enhances in proportion to their frequency when considering variables of successively decreasing subscript $j$. The consequence is the practical absence of higher frequencies in the $z_j$ variables of higher $j$. Instead the variable $\psi$ defined by Eq. (S4) contains equilibrated contributions of the various modes and this makes it a very convenient observable. In addition this variable is sensitive to the relative positions of the fixed points and, then, in the $\psi(t)$ signals (Figs. 2, 3 and S2) the influence of the modes emerged from one fixed point upon those emerged from the other point appear on the top (or bottom) of the waveform oscillations, while the influence among the modes emerged from the same fixed point appear at intermediate positions. This manifests the two basic ways of mode mixing illustrated in Figs. 4A and 4B.

## 2. Difference with the order/disorder of entropy.

The entropic disorder refers to the amorphousness and homogeneity degrees in the spatial distribution of the physical contents of a system, the more homogeneous and amorphous the more disordered, and order simply means lack of disorder. Entropy, denoting the number of possible microstates compatible with the observable macroscopic state, and its associated order/disorder notions are strictly static. They characterize the state of a system at the given moment but not the system in itself and, in particular, do not determine its next future. Such a future depends on the dynamical activity actually occurring within and around the system at the given moment. In those systems where such an activity is dominated by randomness, the second law of thermodynamics applies directly and the system state evolves towards the maximum disorder compatible with the fixed constraints on the system. However, there are systems where the heterogeneities sustain a well-defined ensemble of interactive effects, which is very far from random and from which a peculiar time evolution of the system properties derives. In addition to the entropic order and independently of it, these systems possess what we call dynamical order. To explain where such dynamical order resides and how it can manifest in the system behaviour is the aim of the present article.

Instead we do not consider how the physical structures responsible for it have been originated.

## 3. Descriptive overview of the generalized Landau scenario.

### 3.1. Autonomous systems.

The study of a dynamical system [8,9] implies three kinds of spaces, the less useful of which is the physical space where the system is actually evolving, while the relevant ones are abstract constructs associated with the mathematical description of the given system: the phase space of the system and the space of the dynamical systems. A proper mathematical description of a system within its environment should capture the dynamical effects sustaining the observable time evolution and this means the proper identification of relevant magnitudes and interactive causal influences among them, as well as the proper description of such influences on the time rate of change of the evolving magnitudes. The minimum number $N$ of variable magnitudes required for such a description defines the effective number of degrees of freedom, while the magnitudes remaining fixed in time, usually called parameters, describe properties affecting on but unaffected by the system dynamics and their effects can be interpreted as environmental influences. The constancy of such parameters makes the system dynamically autonomous with respect to the influencing environment and gives sense to analyze its potential time evolutions in the $N$-dimensional phase space defined by the set of variable magnitudes. The constancy of some of such parameters is often a simplifying assumption to avoid the enlargement of the considered system up to include the dynamical reasons of such a parameter variation. This view applies momentarily well when the parameter variations happen independently enough and slowly enough with respect to the system dynamics and, in this way, relatively autonomous subsystems can be temporarily differentiated within larger systems.

The space of dynamical systems is a generic and flexible notion including the whole of conceivable autonomous mathematical systems, a space where every point represents a different system and which is used to compare ones with others in relation to their behaviour, as correspondingly characterized by the overall portrait of phase space trajectories. The neighbourhood of a given system is the Banach space of all its small perturbations but, often, the system is compared with those obtained from it by changing the value of one o more parameters and then the analysis restricts to the consequent parameter space. The primary aim is to find those systems where qualitative changes of behaviour begin to occur in relation to some kind of bifurcation and which define boundary surfaces between regions of behaviourally different systems. The search of systems with richer dynamics and the visualization of the evolutionary pathways towards such systems involve some understanding of such a sort of space. Key points are a) the relative dimension of a given bifurcation surface within the considered space since the higher such a dimension the easier its crossing should be, and this is characterized by the bifurcation codimension that, in practical terms, defines the number of parameters to be simultaneously adjusted in order to catch it, and b) the connections among different kinds of bifurcations since they hierarchically organize the set of boundary surfaces, with the lower codimension bifurcations emanating from those of higher codimension and the latter always implying several lower codimension surfaces getting together. It is seen that expectable evolutionary pathways towards complex dynamics should be based on successions of codimension-one bifurcations, while the higher-codimension ones remain as organizing centres of the bifurcation set.

### 3.2. Oscillations in linear and nonlinear systems.

In the $N$-dimensional phase space of an autonomous system, every point represents a possible state of the system through its coordinates, and their change under the action of the dynamical equations manifests along the time through the trajectory of successive transforming states. The existence and uniqueness theorems assure that the phase space is densely filled of potential trajectories, with a single trajectory passing for each one of its points. The trajectories cannot intersect one another and cannot bifurcate into branches. A typical trajectory represents a set of transient states started at some arbitrarily chosen state and irreversibly evolving towards somewhere, usually one of the so-called limit sets, either fixed points, periodic orbits or others. Such sets are peculiar trajectories that remain within a given region and which are asymptotically connected to other trajectories, either incoming to or departing from them as a function of time. An attracting (repelling) limit set receives (expels) trajectories from (to) all the directions, while a saddle limit set connects with both incoming and departing trajectories in what constitute its stable and unstable invariant manifolds, respectively, as well as there are trajectories first approaching to and then departing from it with bi-asymptotically growing transit times.

Competition among dynamical effects can produce oscillatory evolutions described by trajectories along which the magnitudes of at least two variables alternate increases with decreases under some relative phase difference. They may be either transients or self-sustained oscillations over recurrent trajectories associated with certain kinds of limit sets. These limit sets can appear through two-dimensional bifurcations arising within either the stable or the unstable manifold of a previously existing limit set and the key point is the competitive coordination of two degrees of freedom that become linked in a phase-space zone around the limit set. In this way the two-dimensional process becomes of codimension one and develops under a single parameter variation. There are no other bifurcation processes of dimension higher than one and of codimension one, and this points out the central role of the emergence of self-sustained oscillations in the generation of complex dynamical activity.

The phase space of a linear system generically contains a single limit set, necessarily a fixed point, and its incoming and/or departing trajectories maintain their geometrical and time evolution signatures through the full phase space, i.e., the fixed point neighbourhood defines the full dynamics of the system. The oscillatory possibilities of the linear systems manifest through their single fixed point when, by properly varying parameters, it experiences successive two-dimensional events in different planes up to a maximum number of $N/2$, in each one of which the trajectories become spirals, either incoming to or departing from the fixed point, with a well-defined spiralling frequency. Outside the involved planes, the trajectories describe combinations of the several plane spirals in the form of transient oscillatory evolutions displaying up to $N/2$ different frequencies. The linear systems cannot produce self-sustained oscillations, only transients.

The presence of nonlinearities in the relational dependences of the system allows for new kinds of limit sets, in addition to fixed points, and for the coexistence of a multiplicity of such limit sets in the phase space, with the consequent structure of trajectories asymptotically connecting ones to others. Additional fixed points generically appear by pairs through the saddle-node bifurcation and they emerge with a structure of one-dimensional asymptotic connections among them and to at least one of the previously existing limit sets. All the new kinds of limit sets: the periodic and quasiperiodic orbits and the strange asymptotic sets of chaos, correspond to self-sustained oscillatory time evolutions.

### 3.3. Periodic orbit from a fixed point.

The basic mechanism for generating periodic orbits is the Hopf bifurcation of a fixed point that, like in the linear case, begins with the competing coordination of dynamical effects up to sustain spiral trajectories with a certain frequency in a given two-dimensional surface within one of the invariant manifolds of the point. Unlike the linear case, the spiralling surface now does not extend for the phase space like a plane and, on the other hand, a periodic orbit usually emerges from the fixed point along the two-dimensional surface when the parameter variation originates an exchange of sense in the spiral trajectories from incoming to departing (or vice versa) while the frequency remains unaltered. The orbit is born as an infinitely small circular trajectory around the fixed point by describing a periodic time evolution at the spiral frequency and it maintains the peculiar feature of closing exactly on itself while growing away from the point. It emerges asymptotically connected to the fixed point through the trajectories of the two-dimensional invariant submanifold, whose sense is from the point to the orbit (or vice versa). In the rest of dimensions, the orbit inherits the fixed point stability without altering it and derives its invariant manifolds from those of the point by asymptotically connecting with the same limit sets. This is relevant for realizing how the structure of invariant manifolds develops during the unfolding of the generalized Landau scenario.

While a fixed point describes a state of dynamical equilibrium in which the superposed effects keep all the system variables on fixed values, a periodic orbit describes a self-sustained oscillation affecting the various variables in accordance with the orbit projection upon the respective axes. The shape and orientation of the orbit denote the oscillation amplitudes and relative phases of the different variables, while the period is common for all of them. The oscillation is self-sustained since the sequence of varying dynamical effects associated with the time evolving variables maintain themselves in a recurrent way through the chain of interrelations, and the period just describes the time employed in closing the feedback loop. The oscillatory evolution manifests also outside of the periodic orbit by affecting the neighbouring transient trajectories and, in the case of a saddle orbit, its influence propagates particularly far away along the unstable manifold towards other limit sets by maintaining both the period and phase space orientation of the oscillation.

### 3.4. Mixing of oscillations emerged from different but connected fixed points.

The time evolution of a periodic orbit begins like a harmonic oscillation at the bifurcating frequency but, beyond the bifurcation, the nonlinearities often slow the motion down in a certain orbit zone and consequently alter the time harmonicity and increase the period. Such nonlinear effects are usually associated with the presence of another fixed point, a saddle partner, which, before the Hopf bifurcation, was connected to the bifurcating point through a one-dimensional submanifold that has been now collected by the orbit. It is just along this submanifold that the new orbit develops its growing and manifests the presence of the other fixed point. On the one hand, the slowing down reflects the relative approach of the orbit trajectory to the equilibrium circumstances of that point. On the other hand, if that point has experienced a Hopf bifurcation do not involving the connecting submanifold, the corresponding oscillation mode will propagate along the connection towards the growing orbit and local mixing of oscillation modes will occur in addition to the slowing down effect, as in the case of Fig. 4A. All of these nonlinear effects enhance with the relative proximity of the growing orbit to the external saddle point, up to the limit when the orbit touches and becomes homoclinic to it (or to its periodic orbit), but it is worth noting that the effects can be really significant even far from the homoclinic condition.

The nonlinear mode mixing takes place without requiring any bifurcation of the involved periodic orbits. The influencing orbit simply propagates its oscillations along the unstable manifold and actuates like a corkscrew on neighbouring trajectories and, in particular, on the influenced orbit. This orbit locally transforms by incorporating the varying dynamical effects associated with the influencing oscillation within its own sequence of varying dynamical effects, while the self-sustained adjustment of the feedback loop and the closing on itself feature are maintained without altering the orbit stability. Under the parameter variation, the incorporation process is continuous by beginning from nothing with the gradual appearance of influencing oscillations in the given zone of the orbit. In the time domain, the local mixing manifests as an intermittent burst of influencing oscillations within the cycle of the influenced ones, as well as through the period increase denoting the slowing down effect of the homoclinic approach.

### 3.5. Multiple oscillations from a fixed point and their mixing.

A fixed point can experience successive Hopf bifurcations and originate successive periodic orbits having different frequencies and affecting the various variables differently. The next Hopf bifurcation is not more demanding than the previous one since it will be of codimension-one again and then the sequence of bifurcations requires no other conditions than the sequence of those of the singular bifurcations[1]. In principle, a fixed point can sustain up to $N/2$ bifurcations if both the stable and unstable manifolds are fully engaged but, according what happens in the generalized Landau scenario, we consider bifurcations occurring within the stable manifold only, while the unstable manifold serves to propagate the oscillations towards other limit sets. In addition, for the sake of simplicity, we will consider supercritical bifurcations only. The successive the orbits will emerge with unstable manifolds of successively increasing dimension, since the fixed point will increase in two its unstable dimension at each bifurcation, and, since the point maintains a 2D submanifold connection to every orbit emerged from it, each new orbit will appear with a 3D submanifold of its unstable manifold connected to each one of the previous orbits. It is through such 3D connections that the oscillation modes emerged from the same fixed point mix one another, as shown in Fig. 4B for the simplest case involving the second and first periodic orbits emerged from an initially stable fixed point. Unlike happens in the mixing of oscillation modes emerged from different fixed points, now there are no homoclinic effects and the orbit period does not enlarge when a burst of faster oscillations is incorporated within it. Nonetheless, the nonlinear mode mixing works similarly without requiring any bifurcation of the orbits and the influencing oscillations appear localized on the influenced orbit through the gradual intertwinement produced by the 3D submanifold connection[2].

### 3.6. Quasiperiodic or two-frequency periodic orbits.

The standard way in nonlinear dynamics for generating oscillatory limit sets with two characteristic frequencies is through the torus bifurcation or secondary Hopf bifurcation of a periodic orbit, firstly studied in discrete systems as the Neimark-Sacker bifurcation.

---

[1] The same applies to the Hopf bifurcations of several fixed points. The conditions for a given bifurcation are independent of those of another bifurcation in either the same or different fixed point.

[2] In this case, however, the influencing oscillations can appear in two opposite zones of the influenced orbit [2].

The peculiarities of such a bifurcation with respect to that of a fixed point arise from the fact that the trajectories of both the stable and unstable manifolds of the periodic orbit include indefinite rounds at the orbit frequency in superposition to the attracting and repelling components, respectively[3]. A secondary bifurcation occurring within the stable (unstable) manifold begins again by becoming of codimension-one when the asymptotic approach (departure) incorporates a two-dimensional spiralling component at the secondary frequency and culminates when an exchange of spiral sense originates the appearance of new invariant sets based on trajectories combining the two involved oscillations. The invariant sets include a two-dimensional torus and either one or two limit sets placed on it: a quasiperiodic orbit densely covering the torus surface or a saddle-node pair of periodic orbits asymptotically connected by a two-dimensional submanifold covering the torus surface, and in both cases each limit set appears with the proper structure of invariant submanifolds asymptotically connecting it to the primary orbit and to any other limit set to which that orbit was connected. The occurrence of one or another outcome is when the ratio of the two frequencies is either an irrational or a rational number, respectively. Such a peculiar constriction in the torus oscillatory combination happens because the secondary oscillation is continuously incorporated along each one of the indefinite asymptotic rounds to the bifurcating orbit and this is both the source of the so-called resonance problems of the torus bifurcation and the reason of why the bifurcation occurs with all of its details in the discrete system defined by a Poincaré section of the continuous one. In contrast, it is worth remarking the absence of any resonance problem in the intermittent mode mixing mechanisms, because they incorporate the secondary oscillation in a limited zone of the primary orbit only, as well as the incapability of any Poincaré section to capture the intermittent mode mixing in a periodic orbit.

It is expectable that, similarly to the Hopf bifurcations of a fixed point, a periodic orbit could experience successive torus bifurcations with different secondary frequencies and with the corresponding limit sets asymptotically connected to the primary orbit and among them. There is also the possibility of tertiary and higher-order bifurcations through which an invariant torus of a given order and its limit sets generate a similar structure of higher dimension with an additional frequency, and so on.

---

[3] The attracting and repelling components of the invariant manifolds of a periodic orbit, either stable, unstable or saddle, cover up to $N$-1 dimensions, which are the available ones for the torus bifurcation.

Nevertheless, the invariant tori with their limit sets are delicate structures breaking easily down for a variety of reasons and in a variety of ways, and their existence is consequently limited to rather small regions of the space of dynamical systems [12,13]. Under parameter variation, the invariant torus loses smoothness and vanishes while its limit sets do not disappear into nothing but transform in a complex manner, leading often to the occurrence of chaos [12]. In any case, it is likely that the remnant limit sets would describe time evolutions based on the torus oscillations and that they would maintain asymptotical connection with other limit sets of the basin of attraction.

We are emphasizing the torus bifurcation because the most appropriate circumstance for its occurrence is when a fixed point has done or is near to do a number of Hopf bifurcations, which is just one of the required conditions for the generalized Landau scenario and then the presence of tori or of their remnant limit sets is likely in that scenario[4]. It is worth noting that, under the described circumstance, the oscillations of the two- and multi-dimensional tori are based on those of the Hopf bifurcations of the fixed point in the sense that they have almost equal frequencies and phase space orientations. Thus, such tori do not introduce new oscillation modes but additional mixing mechanisms among the modes of the fixed point. It is also worth remarking the relation between such a kind of torus bifurcation and the intermittent mode mixing between periodic orbits emerged from the same fixed point because both events develop through the same 3D submanifold connecting the two orbits[5].

### 3.7. The generalized Landau scenario.

The optimum development of the generalized Landau scenario [2] implies the coexistence of as much as possible fixed points in the same basin of attraction (one stable at the middle and a cloud of saddles of different unstable dimensions at the basin boundary, with a net of asymptotic connections among them), the occurrence of successive Hopf bifurcations on such fixed points up to exhaust their stable manifolds, possible occurrence of torus bifurcations around each one of the fixed points and its periodic orbits, and the working of the nonlinear mode mixing mechanisms among the variety of coexisting limit sets mutually connected though their structure of invariant manifolds and all of them connected to the attracting set. The generic consequence of such a cumulus of circumstances is essentially twofold: the generation of a multitude of different oscillation modes, each one initially describing a harmonic oscillation characterized by the frequency and the phase-space orientation of the orbit, and the combination of such modes in a variety of mixing pathways affecting all the trajectories in the phase space region where the complex structure of interrelated invariant sets is developing. The mode mixing affects both transient trajectories and limit sets and it is particularly effective upon the attractor since all of the saddle sets have asymptotical connection to it.

---

[4] For instance, the multi-frequency bursts in the time evolution of Fig. 2 could be related to mixing influences of limit sets derived from probably vanished invariant tori.

[5] In some cases, the two kinds of events happen alternatively in one or another of the two orbits. For instance, in the case of Fig. 4B, the orbit 2 will bifurcate subcritically at a higher $\mu_C$ value by becoming stable and originating a saddle torus based on the frequencies of the two orbits while the stable orbit has been locally incorporating the orbit 2 oscillations.

An autonomous system is expected to be evolving according to the time evolution of the attractor, while the multitude of coexisting saddle limit sets are in practice only useful entities for the understanding of the phase-space portrait structure, but every one of the rich variety of transient trajectories asymptotically ending towards the attractor can be effectively induced by displacing the system state with the appropriate external perturbation [2].

**3.8. Difference with chaos.**

It is worth remarking the main difference with respect to chaos, in which a multitude of periodic orbits also coexist in the involved phase space region. The major part of such orbits have appeared through period-doubling, cyclic saddle-node or homoclinic bifurcations, which are one-dimensional events unable to define a new oscillation frequency and which should be more properly considered as producing transformation or destruction rather than creation of characteristic oscillation modes. The numerous periodic orbits are slight variations or combinations of a few basic ones, each one of which is related to a Hopf bifurcation, and the self-sustained sequences of dynamical effects they describe are then similar to those of these basic orbits and their combinations. In other words, the multitude of periodic orbits and the chaotic attractor itself do not imply in practice more dynamical activities than those of the basic orbits. For instance, both the Lorentz and Rössler systems develop their chaotic states over two oscillation modes and the corresponding dynamical activities.

On the other hand, notice that chaos and its complex features can occur in the generalized Landau scenario as a superposed effect. When it happens, every cycle of the chaotic attractor and of each one of the associated non-stable periodic orbits will describe intermittent sequences of all the involved oscillation modes, with irregular cycles in the first case and periodic ones in the second.

### 3.9. Is it possible in discrete dynamical systems?

It is also worth considering the possible occurrence of an equivalent scenario in discrete dynamical systems and the conclusion is that it has no sense because the intermittent mixing will need to occur in between the iterative steps. This is clearly seen in the case of a Poincaré map derived from a continuous system, which provides information on the intersected invariant sets only but not on what has happened to the continuous time evolution between successive intersections. The problem is related to the fact that the intermittent mixing of an arbitrary number of modes can take place in a periodic orbit without requiring its bifurcation and then without requiring the appearance of new invariant sets. In fact, the problem is deeper since the discrete systems cannot experience the primary Hopf bifurcation, which is the true source of oscillation modes in continuous systems, and even less they can describe something equivalent to a variety of fixed points experiencing successive Hopf bifurcations[6].

### 3.10. Final remarks

Finally, notice that the description above corresponds to the behaviour of autonomous systems of ordinary differential equations and that, even under the hypothetic circumstance that the things of the natural world behave like dynamical systems, the autonomy of any physical subsystem under study would be in the best case an approximation and usually it would not apply clearly. This should be taken into account if trying to relate the observed behaviour with the oscillatory scenario. The description corresponds also to clearly dissipative systems since significant differences in the oscillatory scenario take place when the system divergence decreases towards becoming conservative [2]. On the other hand, concerning complexity behaviours, there is the question of to what extent the ordinary differential equations cover all the possible dynamical scenarios or if, conversely, the partial and other kinds of differential equations also employed by the empirical sciences could sustain additional scenarios unachievable with the ordinary ones. It is also in order to consider the variety of discrete systems that, in addition to maps, are being used to numerically illustrate complex behaviours, like cellular automata, coupled map lattices, etc, although their relation with the natural world is less obvious. On the overall this is a broad problem on which we do not enter here.

[6] When dealing with discrete systems, the introduction of an intrinsic periodic evolution by arbitrarily assigning a time unit in between successive iterations has only proper sense in the case of a Poincaré map used to analyse the bifurcations of a periodic orbit of the continuous system, whose period should be the time unit. Even in this case and assuming the simplest circumstance of a harmonic oscillation, the intrinsic period is not enough to describe the associated oscillation since there is a lack of characterization of the relative behaviour of the several variables.